# Non-neutral charged two-dimension system and its quasihole structure

Ting-Ting Kang*
*State Key Laboratory of Infrared Physics, Shanghai Institute of Technical Physics, Chinese Academy of Sciences, 200083 Shanghai, People's Republic of China*
*Corresponding author: ktt219@163.com or kang@mail.sitp.ac.cn

Most of our current knowledge on condensed matters contains a default assumption: the matters are neutral charged. On the other hand, the two-dimension(2D) vortex-Coulomb gas charge analogy is a very successful theoretical tool in explaining superfluid, type-II superconductor and fractional quantum Hall effect(FQHE), because the interaction among vortices show similarity with 2D Coulomb potential. Here, by breaking the "neutral charged" assumption, we suggest the positively charged 2D semiconductor system can possess non-trivial particle-like charge centers(called "quasihole"), which is the charge version of "vortex". Using the hypernetted-chain (HNC) approximation, the structure of quasihole is elucidated. Numerical calculations show that the quasihole can be projected onto another 2D layer, producing an electric field configuration characterized by 3/2 topological charge.

The two-dimensional (2D) Coulomb gas/plasma model, in which, charged particles interact through a Coulomb interaction in a uniform neutralizing background, had played vital roles in modern condensed matter physics [1-6]. For example, vortex in superfluid interacts with each other in a logarithmic potential, which is exactly the same form as 2D Coulomb interaction. Guiding by this 2D vortex-Coulomb gas charge analogy [1], 2D charged particles had been proposed to simulate the vortex in superfluid. This idea was further generalized to type-II superconductor [1-3]. In that case, the vortices generated by an external perpendicular magnetic field [Fig.1(a)] will, for low enough temperature, form a triangular lattice, namely Abrikosov lattice[2,3], which, in the picture of 2D plasma, corresponds to a Wigner lattice by 2D charges of equal sign in a uniform compensating background charge.

In addition to superfluid/superconductor, 2D plasma model continues its surprised success in fractional quantum Hall effect(FQHE)[4,5,6]. A beautiful analogy to 2D plasma was developed by Laughlin regarding FQHE [7,8,9]. According to Laughlin, for a FQHE with filling factor v=1/m (m is an integer), the electron bonded by m quantum flux can be imagined as a particles of charge $me_0$ ($e_0$: charge of a free electron), as depicted by Fig.1(b). But, in this time, FQHE is a quantum liquid, not crystalline. The vortex-charge analogy had even been applied in the field of optics, where the bound states in the continuum are the "vortex" and can be understood in the language of "charge"[10].

The success of 2D plasma model suggests that it may be possible to search a system with "pure" Coulomb interactions to demonstrate similar "vortex" physics in type-II superconductor and FQHE. However, no magnet field is required. Our naive candidate is the 2D non-neutral charged system(2DNNCS), which is necessarily isolated from outside electrically, i.e. ungrounded, like Fig.1(c).

2DNNCS and its transport behaviors, in its own right, is an unexplored and important topic. However, we cannot directly do DC transport measurement by contacting 2DNNCS with electrodes, because any electric contact with the outside will soon turn 2DNNCS back into a neutral charged one. Fortunately, a double quantum well (DQW) scheme, i.e. bilayer, as shown in Fig.1(d), can bypass this dilemma. In this scheme, the DC measurement can done for the low QW (called sensor layer), which is capacitive coupled with the charged object- the upper QW, so the DC information of the charged upper QW can be sent out. This experimental system can be adopted to detect the inhomogeneous electric field produced by 2DNNCS and will be highlighted in this paper.

## Results and Discussions

This paper is divided into 3 parts as following.

**I. The pre-condition for allowing charge non-uniformity.**
Let us consider the configuration in Fig.1(e). For position **r**, if the charges arrangement in upper QW disturbs the electric field from the average value E to E+δE, electrons in upper layer will accumulate at **r** by a density enhancement $\delta N_U$. The net outcome preferred by $\delta N_U$ is the complete smoothing of electric field non-uniformity δE toward zero, which is simply what the screening effect means. So the low layer electron density enhancement $\delta N_L$ will be reduced toward zero. Quantitatively, following the treatment in Ref.(11), we have:

$$\delta\mu_U = \delta\mu_L + e_0 d\delta E \quad (1\text{-}1)$$
$$\delta E = \delta N_L (e_0/\varepsilon_r\varepsilon_0) \quad (1\text{-}2)$$

where $\mu$: chemical potential of electrons; subscript U(L): upper(low) QW; $\varepsilon_r$: dielectric constant of semiconductor; $\varepsilon_0$: vacuum permittivity; $d$: inter-layer distance. Eq.(1-2) is from Gauss's law.

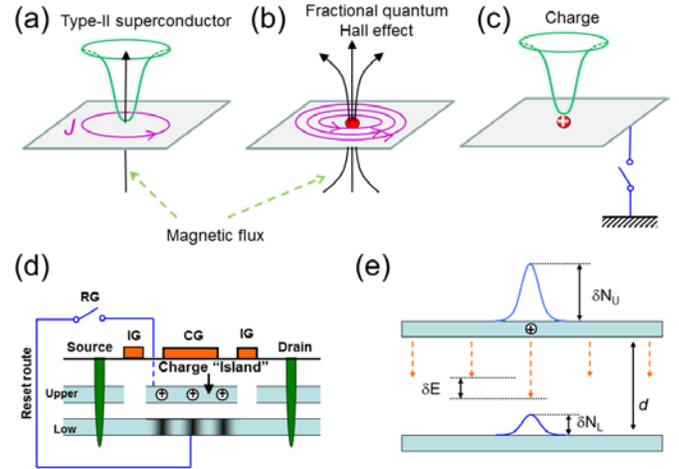

Fig.1 Schematic drawings: (a) a vortex in type-II superconductor, which has a magnetic flux at the center and is surrounded by circulation supercurrents. the superconducting electrons are then expelled from the center, as shown by the green "funnel/ tornado" like plot; (b) v=1/3 fractional quantum Hall effect where 3 fluxes are attached to one electron; (c) A positively charged and ungrounded semiconductor layer. A quasihole is formed by expelling the electrons from the center via a vortex mechanism; (d,e) The GaAs/AlGaAs double quantum well (DQW) based experimental scheme, which is designed to detect the inhomogeneous electric field produced by a charged GaAs QW.

Furthermore, we define the fundamental distance parameter $d_{U(L)} = (\varepsilon_r\varepsilon_0/e_0^2)(\partial\mu/\partial N)_{U(L)}$, which results a convenient relation $\delta\mu_{U(L)} = (\varepsilon_r\varepsilon_0/e_0^2)d_{U(L)}\delta N_{U(L)}$. It notes that $d_{U(L)}$ reflects the compressibility $K$ of QW, because $K^{-1} = N_{U(L)}^2(\partial\mu/\partial N)_{U(L)}$ and can be solved exactly for the case of parabolic two-dimensional electron gas(2DEG), i.e. $d_U = d_L = \pi\varepsilon_r\varepsilon_0\hbar^2/(m^*e_0^2) = a_0/4$ (where $a_0$ is the Bohr radius) [11]. Finally, after simple treatments, we reach:

$$\delta N_L = \delta N_U d_U/(d_L + d) \quad (1\text{-}3)$$

If the upper layer is assumed to be a perfect metal, the metal's infinite large compressibility $K$ (which means $\partial\mu/\partial N = 0$) immediately leads to $d_U = 0$. In this case, although $\delta N_U$ is not zero, we still reach $\delta N_L = 0$, δE=0 (by Gauss's law), making the electric field trivially uniform. Fortunately, semiconductor 2DEG doesn't have perfect screening ability. Concerning GaAs 2DEG, $d_U = a_0/4 \approx 2.5$nm can be non-zero. This small but non-zero $\delta N_L$ (or $a_0$) value makes semiconductor (rather than metal) based 2DNNCS special.

**II. The HNC treatment of a quasihole.**

The simple method above cannot address the detailed structure of "quasihole", because they are only a large-scale description. To fill this gap, we use the hypernetted-chain (HNC) approximation[7-9, 12-13] to calculate for the electron density distribution, i.e. the radial distribution function $g(r)$, inside/around a quasihole. HNC treats a classical liquid with a short-range pair potential $u(r)$ and constant external potential. The attractiveness of HNC is that if $u(r)$ is known, $g(r)$ can be exactly calculated. Although HNC is originally not a quantum method, its ability in classical mapping quantum objects had been widely recognized [14].

A two-component HNC is adopted [14]:

$$h_{12}(q) = c_{12}(q)[1 + \rho_1 h_{11}(q)] \quad (2\text{-}1)$$
$$g_{12}(r) = \exp[N_{12}^s(r) - u_{12}^s(r)] \quad (2\text{-}2)$$
$$h_{12}(r) = g_{12}(r) - 1 \quad (2\text{-}3)$$

where the index 1 denotes a normal electron, and 2 denotes a phantom particle(i.e. quasihole) with the charge quantity of $m_2 e_0$, $\rho_1$ is the number density of species 1. In the above formula, the considered system is a quasihole at the origin and essentially no other quasiholes in the system. It therefore suffices to set $\rho_1 = 1$. The electron density surrounding the quasihole is then described by $n(r) = n_0 g_{12}(r)$, where $n_0$ is the electron density for the QW which hosting the quasihole. The Fourier transform of between radius $r$ and wave vector $q$ is:

$$F(r) = \int_0^\infty F(k)J_0(qr)qdq/2\pi, \quad F(q) = 2\pi\int_0^\infty F(r)J_0(qr)rdr \quad [15],$$

where $J_0$ is the zeroth-order Bessel function.

To work with the long-range forces in plasmas, the HNC equations should be regularized, which considers the coulomb potentials to be the limits of short-range potentials. The practical doing is breaking up $u(r)$ into long- and short-range pieces: $u(r) = u^s(r) + u^l(r)$. And the short-range correlation functions in Eq.(2) are defined as:

$$N_{12}^s(q) = h_{12}(q) - c_{12}^s(q) \quad (3\text{-}1)$$
$$c_{12}(q) = c_{12}^s(q) - u_{12}^l(q) \quad (3\text{-}2)$$

For our HNC calculation, we have chosen the length unit 1 to be $r_{HNC}$, where $r_{HNC} = (2/n_0)^{1/2}$. On $u(r)$, we start from the 2D Poisson's equation for a 2DEG subjected to a point charge with charge quantity $+e_0$ at **r**=0 ( a 2D coordinate):

$$\nabla^2 U(\mathbf{r}) - \frac{U(\mathbf{r})}{l_0^2} = \frac{n_{ext}}{\varepsilon_r\varepsilon_0}e_0 = \frac{\delta(\mathbf{r})}{\varepsilon_r\varepsilon_0\delta_z}e_0 \quad (4)$$

where $U$ is the electric potential, $l_0$ is the screening length of 2DEG and calculated as $l_0^2 = \varepsilon_r\varepsilon_0\delta_z/D = a_0\delta_z/4$, $\delta_z$ is characteristic thickness of the QW, which meets $\delta_z \sim a_0$. A reasonable approximation may be $l_0 \approx a_0$. Eq.(4) has an analytical solution $U(r) = K_0(r/l_0)e_0/(\pi\varepsilon_r\varepsilon_0\delta_z)$, where $K_0$ is the modified Bessel function of the second kind. This indicates that $u_{12}$ can be written as $u_{12} = 2m_2 K_0(q_c r)$, with $q_c = 1/l_0$.

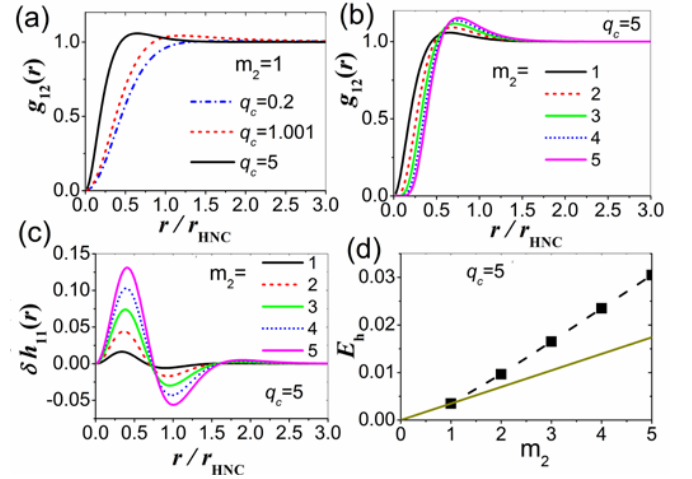

Fig.2. For the quasihole: (a) $g_{12}(r)$ under different $q_c$; (b) $g_{12}(r)$ under different $m_2$; (c) $\delta h_{11}(r)$ under different $m_2$; (d) $E_h$ vs $m_2$.

Following Laughlin's treatment [9], we define:

$$u_{12}^l(q) = 4\pi m_2 \frac{Q^2}{(q^2 + q_c^2)(q^2 + Q^2)} \quad (5)$$

where $Q$ is a "cut-off" momentum in the order of 1 and the obtained $g(r)$ doesn't depend on $Q$ [8,12]. So, by Fourier transform, we reach:

$$u_{12}^l(r) = \int u_{12}^l(q) e^{i\vec{q}\cdot\vec{r}} d^2q/(2\pi)^2$$
$$= 2m_2 \frac{Q^2}{(Q^2 - q_c^2)}[K_0(q_c r) - K_0(Qr)] \quad (6)$$

With $u_{12}$ and $u_{12}^l$, we still need $g_{11}$ [which is required to calculate $h_{11}(q)$] to calculate $g_{12}$. The pair correlation function for electrons is analytically known as $g_{11}(r)=1-\exp(-\pi r^2)$, which describes the pair correlation function for electrons in 2D classical one-component plasma (2D-OCP) [16]. v=1 FQHE also has the similar $\sim -\exp(-br^2)$ form with $b$ being a different constant [17], because its analogy with 2D-OCP.

With the preparations above, we are ready to use HNC to calculate the spatial electron distribution $g_{12}(r)$ of a quasihole. In Fig.2(a), we set the quasihole's charge quantity to be one elementary charge, i.e. $m_2=1$, and study the influence of $q_c$ on $g_{12}(r)$. The $q_c$ value reflects the ability of the electrons to smoothen away the quasihole: $q_c \to \infty$ corresponds to metal and $q_c \to 0$ corresponds to insulator (where no mobile carriers is available for screening). The intention of Fig.2(a) is to clarify that how the two length scale, i.e. screening length $l_0$ and $r_{HNC} \propto n_0^{-1/2}$, cooperates together to shape $g_{12}(r)$.

Generally, in Fig.2(a), the quasihole can be described by a core-shell structure: (i) core region, where $g_{12}(r) \sim 0$ and $g_{12}(r)<1$; (ii) shell region, where $g_{12}(r)>1$ and $g_{12}(r) \sim 1$ in the long range. When $q_c$ is reduced, e.g. from 1.001 to 0.2 in Fig.2(a), the core region will expand. But the $g_{12}(r) \sim 0$ region will not be infinitely expanded and its ultra-limit line-shape at $q_c=0$ is exactly $\lim_{q_c \to 0} g_{12}(r) = 1 - \exp(-\pi r^2)$, i.e. the same as $g_{11}(r)$.

On the other hand, if we enhance the QW's metallicity by increasing $q_c$, e.g. from 1.001 to 5 in Fig.2(a), the size of core will decrease as expected. The ultra-limit of this metallic trend is the complete removal of the $g_{12}(r) \sim 0$ core and resulting $g_{12}(r)=1$ everywhere. This can be easily verified by setting $q_c=\infty$ [so $u_{12}^l(q)=0$, $u_{12}(q)=0$] in the HNC calculation, and satisfactorily explains why a 2D metal system can not support the existence of non-uniform electric field, no matter it is charged or not, as we have reiterated in part I.

There is another fundamental question left, i.e. how many elementary charges are carried by one quasihole (or the "$m_2=$?" question). In other words, if the 2DNNCS has $+N_e e_0$ charges, it prefers $N_e$ quasiholes(each has $+e_0$ charge and $N_e=2,3,4…$) or one quasihole with $+N_e e_0$ charges, or other allocation ways? This can be clarified by comparing the creation energy $\xi_h$ of quasihole with different $m_2$.

The energy $\xi_h$ to make a quasihole is proportional to the integral $E_h$: $\xi_h \propto \int_0^\infty \delta h_{11}(r) dr = E_h$, where $\delta h_{11}(r)$ is the change in $h_{11}(r)$ due to the presence of the quasihole [12,13]. $\delta h_{11}(r)$ can be calculated by the standard procedures [12, 13, 18]. Fig.2 (b,c) present $g_{12}(r)$, $\delta h_{11}(r)$ with varying $m_2$ value. Consequently, in Fig.2(d), we find the relation $E_h(m_2=1)<E_h(m_2>1)/m_2$ with $m_2=2,3,4,5…$, which means: creating $N_e$ quasiholes(each has $+e_0$ charge) needs less energy than producing one quasihole with $+N_e e_0$ charges, namely $m_2=1$ is the favorite choice in respect of energy.

**III. Electric field distribution and its topological structure.**

Following the pseudospin language for bilayer [19-25], in which pseudospin up means that the electron is in the upper layer and pseudospin down means that the electron is in the low layer, we can define the order parameter in our non-neutral charged bilayer to be [19]: $\Psi(\mathbf{r})=|\Psi(\mathbf{r})|\exp[i\varphi(\mathbf{r})]=<\psi_U^+(\mathbf{r})\psi_L(\mathbf{r})>$, where $\psi^+$ and $\psi$ are electron creation and annihilation operators, $\varphi(\mathbf{r})$ is the condensate phase. This order parameter $\Psi(\mathbf{r})$ physically represents an electron in low layer bound to a hole in the upper layer. The order parameter field is then the local charge, or local pseudospin magnetization.

For the low layer, since the charge density $\rho$ is proportional to the local electric field $\mathbf{E}$ via the Gauss's law $\rho=\varepsilon_r\varepsilon_0|\mathbf{E}|$ [we ignore the influence of $\mathbf{E}$ field penetration, because this penetrated electric field is a small term compared with the $\mathbf{E}$ value just above low layer, as depicted in Fig.3(a)], the local order parameter orientation is then decided by the electric field direction. We define the order in terms of a local unit vector field $\mathbf{m}(\mathbf{r})$[22,23]:

$$\mathbf{m}(\mathbf{r})=\Psi(\mathbf{r})/|\Psi(\mathbf{r})|\approx\mathbf{E}(\mathbf{r})/|\mathbf{E}(\mathbf{r})|$$

This order parameter is nonzero only when phase coherence is established between the layers. i.e. the electron in low layer is subjected to a non-zero electric field.

To calculate the electric field spatial distribution produced by a quasihole in this DQW configuration, we have developed a method within the framework of multi-component random phase approximation (RPA) [18, 26]. For our calculation concerning a GaAs/AlGaAs DQW, in a cylindrical coordinate shown in Fig.3(a)-right, a quasihole is placed at R=0, z=d=100nm and the low QW is at z=0 plane. We choose the electron density in upper (low) QW to be $n_U$=4.5×10$^{11}$ cm$^{-2}$($n_L$=3.5×10$^{11}$ cm$^{-2}$), which are the typical values in experiments.

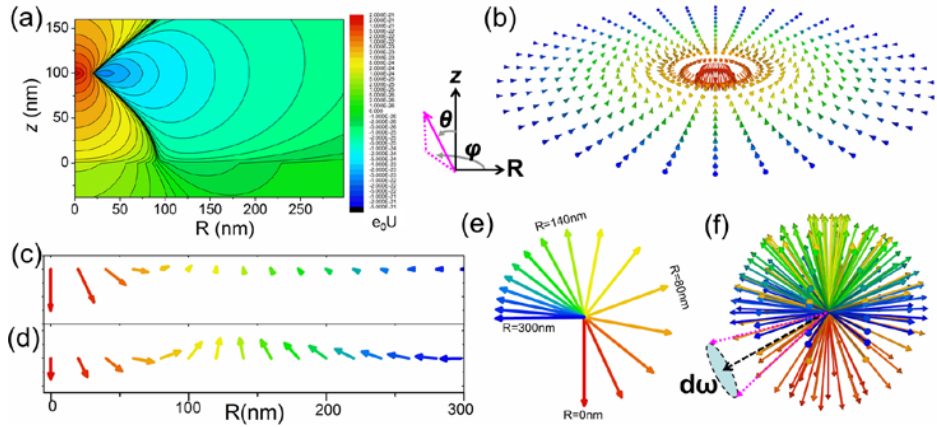

Fig.3 For a GaAs/AlGaAs double quantum well(QW) structures with upper/low QW electron density $n_U=4.5\times10^{11}cm^{-2}$/ $n_L=3.5\times10^{11}cm^{-2}$ (so $q_c\approx2.12$), inter-layer distance $d=100nm$. (a) the spatial distribution of electric potential. (b-f)At the low layer plane ,i.e. z=0: (b) a 3D plot of the electric field $\mathbf{E}(\mathbf{r})$ distribution.(c) electric field $\mathbf{E}$ vs R. (d) $\mathbf{m}$ (the unit vector representing the local electric field orientation) vs R. (e, f) the sweeping trajectory of $\mathbf{m}(\mathbf{r})$ in the unit sphere. In (b,c,d,e,f), the arrow's start point is at the position $\mathbf{r}$. The arrow's direction indicates the direction of the $\mathbf{E}$ (or $\mathbf{m}$) at position $\mathbf{r}$, the length of the line behind the arrow indicates the amplitude of $\mathbf{E}$ (or $\mathbf{m}$) at position $\mathbf{r}$. We use different color to label the arrows at different R position: with positon varying from R=0 to R=300nm, the color changes from red to blue.

Fig.3(a) shows the electric potential in the whole space. We can clearly see that how the core(positively charged) and shell(negatively charged) component of quasihole competes with each other, and roughly control the near and far field pattern respectively. The electric field which penetrating the low GaAs layer is also clearly shown. Anyway, an electromagnetic computation method based on quantum mechanics is indispensable [26].

The texture-like electric field **E** distribution is then straightforwardly reached by differentiating the electric potential. Fig.3(b) shows the electric field distribution at z=0 plane(low layer), with arrow's direction(the length behind arrow) depicting the direction(magnitude) of **E(r)** at the spatial position **r**=(R, φ) with R(φ) being the radius(azimuthal angle). On the in-plane **E(r)** distribution, according to the topological classification, it corresponds to a "vortex/skyrmion" configuration with +1 vorticity, phase γ=0 [27]. The cross sectional view of **E**[Fig.3(c)] and **m**[Fig.3(d)] along the radius is provided for a clear view. We find: **E** is down at the origin and gradually turns up at intermediate distances R~d, finally lies in the *x-y* plane at infinite radius.

The **m(r)** in 2D *x-y* plane of low layer may be mapped to a unit sphere with all start points at the sphere center, as shown in Fig.3(f). Consequently, the electric filed texture configuration is characterized by the topological charge $Q_{top}$ defined by[20,27]:

$$Q_{top} = \iint \rho_{top} dxdy = \frac{1}{4\pi} \iint \mathbf{m} \cdot (\frac{\partial \mathbf{m}}{\partial x} \times \frac{\partial \mathbf{m}}{\partial y}) dxdy \quad (7)$$

as the integral of the solid angle $d\omega = \mathbf{m} \cdot (\frac{\partial \mathbf{m}}{\partial x} \times \frac{\partial \mathbf{m}}{\partial y}) dxdy$, and counts how many times **m(r)** = **m**(x,y) wraps the unit sphere, as shown in Fig.3(f). $\rho_{top}$ is then the topological charge density.

As presented in Fig.3(e,f), a 6π solid angle has been swept from R=0 to R=∞, so this **E** texture carries 3/2 topological charge, which differs from some well-known topological excitations, like skyrmions($Q_{top}$=1) [22,28] and merons ($Q_{top}$=1/2) [22]. $Q_{top}$ is a topological invariant, which is stable against smooth continuous distortions of the field **m**. For example an electron density plasma wave could pass through the texture and $Q_{top}$ would remain invariant.

## Summary

We establish the theoretical framework to treat the behaviors of non-neutral(positive) charged two-dimension system. Its central foundation is the Coulomb version of hypernetted-chain approximation. This makes it possible to calculate the structure of quasihole, on which everything in this work is based. Different from the quasiparticles in fractional quantum Hall effect, where only one length scale(i.e. the quantized cyclotron orbit radius) is related, the charge quasihole's structure is decided by two length scales, i.e. the Bohr radius and the average inter-electron distance.

The vortex-charge analogy reflects a common strategy among scientists: using a simple concept ("charge") to understand a complicated physics ("vortex"). Therefore, the "charge" concept is just a theoretical tool, which is too simple and doesn't deserve further studies. This might be the reason why vortex-charge analogy was independently proposed in different fields [1,7,29], but "charge" itself attracts much less interests. Importantly, our work clarifies that the behavior of experimental "charge" (quasihole) largely agrees with the topological "vortex" excitation regarding electron pseudospin distribution. We still need more experiments to know that this "vortex" behavior will exert what impacts on quantum Hall effect and superfluid/superconductivity, etc. Anyway, non-neutral charged two-dimension system will bring us a fresh experimental tool-box and scientific perspective.

Acknowledgement .- The work was supported by the Natural Science Foundation of China(NSFC) 11204334, 61475178, the "Hundred Talent program" of the Chinese Academy of Sciences. The supports from NSFC 61574150, 61376015, 91321311, Shanghai Science and Technology Foundation 14JC1406600 are also acknowledged.

[1] Petter Minnhagen, Rev. Mod. Phys. 59, 1001 (1987). The two-dimensional Coulomb gas, vortex unbinding, and superfluid-superconducting films.
[2] A. A. Abrikosov, Rev. Mod. Phys. 76, 975 (2004). Nobel Lecture: Type-II superconductors and the vortex lattice.
[3] David A. Huse , Matthew P. A. Fisher and Daniel S. Fisher, Nature 358, 553 (1992). Are superconductors really superconducting.
[4] R. B. Laughlin, Rev. Mod. Phys. 71, 863 (1999). Nobel Lecture: Fractional quantization.
[5] Jainendra K. Jain, Physics Today 53(4), 39 (2000). The Composite Fermion: A Quantum Particle and Its Quantum Fluids.
[6] Yasuhiro Iye, Proc. Natl. Acad. Sci. USA 96, 8821 (1999). Composite fermions and bosons: An invitation to electron masquerade in Quantum Hall.
[7] R. B. Laughlin, Phys. Rev. Lett. 50, 1395 (1983). Anomalous Quantum Hall Effect: An Incompressible Quantum Fluid with Fractionally Charged Excitations.
[8] R.B. Laughlin, Surface Science 142, 163 (1984). Primitive and composite ground states in the fractional quantum hall effect.
[9] R. B. Laughlin, in *The Quantum Hall Effect*, Richard E. Prange and Steven M. Girvin eds., Springer-Verlag, New York (1987).
[10] Bo Zhen, Chia Wei Hsu, Ling Lu, A. Douglas Stone, and Marin Soljačić, Phys. Rev. Lett. 113, 257401(2014). Topological Nature of Optical Bound States in the Continuum.
[11] J. P. Eisenstein, L. N. Pfeiffer, and K. W. West, Phys. Rev. B 50, 1760 (1994). Compressibility of the two-dimensional electron gas: Measurements of the zero-field exchange energy and fractional quantum Hall gap.
[12] H. A. Fertig and B. I. Halperin, Phys. Rev. B 36, 6302 (1987). Hypernetted-chain approximation and quasiparticle energies in the (1/3 fractional quantized Hall effect.
[13] Tapash Chakraborty, Phys. Rev. B 31, 4026 (1985). Elementary excitations in the fractional quantum Hall effect.
[14] François Perrot and M. W. C. Dharma-wardana, Phys. Rev. Lett. 87, 206404 (2001). 2D Electron Gas at Arbitrary Spin Polarizations and Coupling Strengths: Exchange-Correlation Energies, Distribution Functions, and Spin-Polarized Phases.

[15] Béal-Monod, M. T, Phys. Rev. B 36, 8835 (1987). Ruderman-Kittel-Kasuya-Yosida indirect interaction in two dimensions.
[16] B. Jancovici, Phys. Rev. Lett. 46, 386 (1981). Exact Results for the Two-Dimensional One-Component Plasma.
[17] R. K. Kamilla, J. K. Jain, and S. M. Girvin, Phys. Rev. B 56, 12411 (1997). Fermi-sea-like correlations in a partially filled Landau level.
[18] See the APPENDIX.
[19] Jung-Jung Su and A. H. MacDonald, Nature Phys. 4, 799 (2008). How to make a bilayer exciton condensate flow.
[20] S.M. Girvin, in *Topological Aspects of Low Dimensional Systems*, NATO ASI, Les Houches Summer School, A. Comtet, T. Jolicoeur, S. Ouvry and F. David Eds, Springer, p.551–570 (1999).
[21] S.M. Girvin, in *Perspectives in Quantum Hall Effects*, Edited by Sankar Das Sarma and Aron Pinczuk (Wiley, New York, 1997).
[22] K. Moon, H. Mori, Kun Yang, S. M. Girvin, A. H. MacDonald, L. Zheng, D. Yoshioka, and Shou-Cheng Zhang, Phys. Rev. B 51, 5138 (1995). Spontaneous interlayer coherence in double-layer quantum Hall systems: Charged vortices and Kosterlitz-Thouless phase transitions.
[23] Kun Yang, K. Moon, Lotfi Belkhir, H. Mori, S. M. Girvin, A. H. MacDonald, L. Zheng, and D. Yoshioka, Phys. Rev. B 54, 11644 (1996). Spontaneous interlayer coherence in double-layer quantum Hall systems: Symmetry-breaking interactions, in-plane fields, and phase solitons.
[24] J. P. Eisenstein and A. H. MacDonald, Nature 432, 691–694 (2004). Bose-Einstein condensation of excitons in bilayer electron systems.
[25] Steven M. Girvin, Physics Today 53(6), 39 (2000). Spin and Isospin: Exotic Order in Quantum Hall Ferromagnets.
[26] Ting-Ting Kang, arXiv:1303.1067. (2013). Screening in coupled low-dimensional systems: an effective polarizability picture.
[27] Naoto Nagaosa and Yoshinori Tokura, Nature Nanotech. 8, 899 (2013). Topological properties and dynamics of magnetic skyrmions.
[28] Jiadong Zang, Maxim Mostovoy, Jung Hoon Han, and Naoto Nagaosa, Phys. Rev. Lett. 107, 136804 (2011). Dynamics of Skyrmion Crystals in Metallic Thin Films.
[29] K. K. Mon and S. Teitel, Phys. Rev. Lett. 62, 673 (1989). Phase Coherence and Nonequilibrium Behavior in Josephson Junction Arrays

## APPENDIX

### (1) Derivation of $\delta N_L$, $\delta N_U$ in Part I.

The derivation is based on the method by J. P. Eisenstein *et.al* [Compressibility of the two-dimensional electron gas: Measurements of the zero-field exchange energy and fractional quantum Hall gap, *Phys. Rev. B* **50**, 1760 (1994).]

We start from

$$\begin{cases} \delta\mu_U = \delta\mu_L + e_0 d\delta E \\ \delta E = \dfrac{e_0}{\varepsilon_r \varepsilon_0}\delta N_L \end{cases} \quad (S1\text{-}1)$$

The second equation is from Gauss's law.

So $\delta\mu_U = \delta\mu_L + e_0 d\delta E = \delta\mu_L + \dfrac{e_0^2}{\varepsilon_r \varepsilon_0} d\delta N_L$ (S1-2)

We also have the fundamental distance parameters

$$d_{U(L)} = \frac{\varepsilon_r \varepsilon_0}{e^2}\left(\frac{\partial \mu}{\partial N}\right)_{U(L)} \quad (S1\text{-}3)$$

Then (Eq. S1-2) can be processed as below:

$$\delta\mu_U = \delta\mu_L + e_0 d\delta E = \delta\mu_L + \frac{e_0^2}{\varepsilon_r \varepsilon_0} d\delta N_L \Rightarrow$$

$$\frac{e_0^2}{\varepsilon_r \varepsilon_0} d_U \delta N_U = \delta\mu_U = \delta\mu_L + \frac{e_0^2}{\varepsilon_r \varepsilon_0} d\delta N_L = \frac{e_0^2}{\varepsilon_r \varepsilon_0} d_L \delta N_L + \frac{e_0^2}{\varepsilon_r \varepsilon_0} d\delta N_L$$

$$\Rightarrow d_U \delta N_U = d_L \delta N_L + d\delta N_L$$

$$\Rightarrow \delta N_L = \frac{d_U}{d_L + d}\delta N_U$$

Finally we reach $\delta N_L = \dfrac{d_T}{d_L + d}\delta N_U$. (S1-4)

### (2) The calculation of $\delta h_{11}(r)$.

$\delta h_{11}(r)$ can be calculated by the following equations:
$c_{12}(q) = h_{12}(q)/[1 + h_{11}(q)]$
$c_{11}(q) = h_{11}(q)/[1 + h_{11}(q)]$

$\delta N_{11}^s(q) = \{\delta c_{11}^s(q)[h_{11}(q)+c_{11}(q)]+c_{12}(q)h_{12}(q)\}/[1-c_{11}(q)]$

$\delta c_{11}^s(r) = \delta N_{11}^s(r)[g_{11}(r)-1]$

$\delta h_{11}(r) = \delta N_{11}^s(r)g_{11}(r)$

### (3) The calculation of electric potential and electric field.

The calculation of spatial electric potential $U(r,z)$ produced by a quasihole (with its center at $z=d=100nm$, $r=0$) in upper QW is based on the following equations:

$$\rho_{hole}(k) = \int_0^\infty \rho_{hole}(r)J_0(kr)rdr \qquad \text{(S3-1)}$$

$$U(r,z) = \int_0^\infty \rho_{hole}(k)J_0(kr)V(k,z)kdk \qquad \text{(S3-2)}$$

$$V(k,z) = V_0(k)(e^{-k|z-d|} + \frac{V_{LLLL}\Pi_{LL}}{1-V_{LLLL}\Pi_{LL}}e^{-k(d+|z|)}) \qquad \text{(S3-3)}$$

where $\rho_{hole}(r)$ is the positive charge density distribution of the quasihole; $r$ is the radius; subscript U(L) : upper(low) layer index; $V_0(k,z) = V_{LLLL} = \frac{e_0^2}{2\varepsilon_r\varepsilon_0 k}$ is the 2D Fourier transform of bare coulomb interaction. $\Pi_{LL}$ is the 2D educible non-interacting electronic polarizability function of the low layer.

The derivation of Eq.(S3-2, S3-3) is based on a multi-component random phase approximation(RPA) method, which had been detailed explained in our previous paper[ Ting-Ting Kang, *arXiv*:1303.1067. (2013). Screening in coupled low-dimensional systems: an effective polarizability picture.] .